\documentclass[twocolumn,aps,prb,amsmath,showpacs]{revtex4}
\usepackage[dvips]{graphics}
\newcommand{\be}{\begin{equation}}
\newcommand{\ee}{\end{equation}}

\newcommand{\bea}{\begin{eqnarray}}
\newcommand{\eea}{\end{eqnarray}}
\newcommand{\bd}{\begin{displaymath}}
\newcommand{\ed}{\end{displaymath}}
\newcommand{\bi}{\begin{itemize}}
\newcommand{\ei}{\end{itemize}}
\newcommand{\bc}{\begin{center}}
\newcommand{\ec}{\end{center}}
\newcommand{\bfl}{\begin{flushleft}}
\newcommand{\efl}{\end{flushleft}}
\newcommand{\bfr}{\begin{flushright}}
\newcommand{\efr}{\end{flushright}}
\newcommand{\f}{\frac}

\def\br{{\bf r}}
\def\bk{{\bf k}} \def\bq{{\bf q}}

\def\ra{\rightarrow}
\def\6{\partial} \def\a{\alpha} \def\b{\beta}
 \def\d{\delta} \def\ve{\varepsilon}

\def\ss{\sigma} 

\def\o{\omega}

\def\={\!\!\!&=&\!\!\!}
\def\+{\!\!\!&&\!\!\!+~}
\def\-{\!\!\!&&\!\!\!-~}

\begin{document}
\title{Fluctuation conductivity in layered $d$-wave superconductors near critical disorder}
\author{I. Tifrea$^{1,\; 2}$, D. Bodea$^{1,\; 3}$, I. Grosu$^1$, and M.
Crisan$^{1}$}

\affiliation{$^1$Department of Theoretical Physics, University of
Cluj, 3400 Cluj-Napoca, Romania}

\affiliation{$^2$Department of Physics and Astronomy, University
of Iowa, Iowa City, IA 52245, USA}

\affiliation{$^3$Max Plank Institute for the Physics of Complex
Systems, 01187 Dresden, Germany}

\begin{abstract}
We consider the fluctuation conductivity in the critical region of
a disorder induced quantum phase transition in layered $d$-wave
superconductors. We specifically address the fluctuation
contribution to the system's conductivity in the limit of large
(quasi-two-dimensional system) and small (quasi-three-dimensional
system) separation between adjacent layers of the system. Both
in-plane and $c$-axis conductivities were discussed near the point
of insulator-superconductor phase transition. The value of the
dynamical critical exponent, $z=2$, permits a perturbative
treatment of this quantum phase transition under the
renormalization group approach. We discuss our results for the
system conductivities in the critical region as function of
temperature and disorder.
\end{abstract}
\date{\today}
\maketitle

\section{Introduction}

A phase transition at $T=0$ is usually addressed as a quantum
phase transition (QPT).\cite{sachdev} In general, QPT's are driven
by quantum fluctuations controlled by a non-thermal parameter,
namely by impurities, pressure or magnetic fields.\cite{baron} In
the case of high temperature superconductors (HTSC) a phase
transition can be driven both by disorder or
doping.\cite{fukuzumi, karpinska} As function of disorder the
transition is observed for magnetic and nonmagnetic impurities,
when the impurity concentration is high enough to destroy the
phase coherence in the system. As function of doping the standard
phase diagram of HTSC presents two possible QPT's, corresponding
to the end points of the superconducting region, one in the
underdoped region and the other one in the overdoped
region.\cite{schneider1,schneider} In the underdoped limit the
transition is of superconductor-insulator type,\cite{fukuzumi}
whereas in the overdoped limit the transition is of
superconductor-normal state (metal) type.\cite{momono}

Modelling high temperature superconductors (HTSC) as two
dimensional (2D) systems can be justified by the presence of
CuO$_2$ planes in their structure. However, when transport
properties in the critical region around the superconducting phase
transition are considered, the theoretical approach has to take
into account the quasi-2D nature of the system, as it is known
that an important contribution to the conductivity is given by
unusual strong fluctuations. Different from metallic
superconductors (MS), HTSC present a strong anisotropy, leading to
differences in the temperature dependance of the transverse and
in-plane resistivities, a property which is hard to explain based
on the conventional theory of Fermi liquids.\cite{anderson} With
this in mind, the simplest way to include the third dimension of
the system ($c$-axis) is to consider a layered structure in which
adjacent CuO$_2$ are coupled through a tunnelling like term. In
the presence of disorder the role of fluctuations increases, their
effect on the transport properties being very important even for
the case of MS.\cite{larkin} In HTSC, fluctuations are amplified
respect to the MS case, their effects being of main importance as
transport along $c$-axis is considered,\cite{varlamov} scattering
by virtual Cooper pairs increasing the transverse resistivity, in
contrast with the usual effect observed in isotropic MS. The
analysis of the fluctuations effects in superconducting materials
can not distinguish the general symmetry of the order parameter,
which is of $s$-wave type in MS and of $d$-wave type in
HTSC.\cite{scalapino,tsuei} The differences between the two types
of symmetry reduce to a constant numerical prefactor, which from
the experimental point of view is hard to examine. A more
important role is played by the system
dimensionality.\cite{skocpol} However, the analysis of the
fluctuation conductivity in systems with a $p$-wave symmetry of
the order parameter such as Sr$_2$RuO$_4$,\cite{maeno} reveals the
possibility of tracing the pair symmetry in such
systems.\cite{adachi}

QPT's are fundamentally different from finite temperature phase
transitions as a dynamical critical exponent, $z$, needs to be
considered in order to apply the scaling theory to quantum
criticality.\cite{hertz} For the case of HTSC the idea of quantum
criticality was largely explored, indications of a 2D
superconductor to insulator phase transition with a $z=1$ critical
exponent in the underdoped regime, and a three dimensional (3D)
superconductor to normal state phase transition with $z=2$ in the
overdoped regime being identified.\cite{schneider1,schneider} The
disorder induced QPT in $d$-wave HTSC was analyzed both at
$T=0$\cite{herbut} and at finite temperature,\cite{crisan}
considering a two dimensional (2D) system and a dynamical critical
exponent $z=2$. For the 2D case, a detailed discussion of the
in-plane conductivity was done by Herbut\cite{herbut} at $T=0$ and
by Dalidovich and Phillips,\cite{phillips1,phillips2} with the
main conclusion that the dc conductivity will have a {\it
nonuniversal} singular part at the transition point in the quantum
critical regime.

In this work we will analyze the disorder induced QPT in 3D
$d$-wave HTSC considering a layered structure of the system. Such
an analysis will give us the possibility to study the dimensional
crossover of the system as function of the interlayer distance
($s$). Despite the fact that a layered system is by definition an
anisotropic 3D system, at large interlayer distances ($s\ra
\infty$) the system approaches a quasi 2D-structure, such that by
varying $s$ between 0 and $\infty$ we basically interpolate
between 3D and 2D. In the critical region fluctuation effects
dominate the in-plane and c-axis conductivities, which in our
calculation are considered of the Aslamazov-Larkin\cite{larkin}
form. On the other hand, for a quasi-3D system, at small
separation between adjacent layers, the conductivity conserve the
nonuniversal behavior in the quantum critical regime observed in
the 2D case.\cite{phillips2}

The paper is organized as follows. Section II presents our model
for the fluctuation propagator in the presence of disorder for the
case of a layered $d$-wave superconductor. Section III presents
the general equations of the renormalization group approach along
with their solutions for two particular regimes, namely, the
quantum disorder (QD) and quantum critical (QC) regimes. Section
IV presents analytical solutions for the in-plane and $c$-axis
conductivities for the case of large and small interlayer
separation in both QD and QC regimes. Finally, we give our
conclusions.

\section{Fluctuation propagator in layered systems}

We consider that a layered $d$-wave superconductor in the presence
of disorder can be described by a BCS type Hamiltonian with an
additional term corresponding to a random interaction potential
\bea\label{hamiltonian}
&&{\cal
H}=\sum_{\bk,\ss}\ve(\bk)a^\dagger_\ss(\bk)a_\ss(\bk)\nonumber\\
&&-\f{1}{2}\sum_{\ss,\ss'}\sum_{\bk,\bk',\bq}
V(\bk,\bk')a^\dagger_\ss(\bk_+)a^\dagger_{\ss'}(-\bk_-)a_{\ss'}(-\bk'_-)a_\ss(\bk'_+)
\nonumber\\
&&+\sum_{i}\int d\br_{\parallel}
U_i(\br_\parallel)a^\dagger_\ss(\br_\parallel,i)a_\ss(\br_\parallel,i)\;,
\eea
where $\bk_\pm=\bk\pm\bq/2$ and $a_\ss(\br_\parallel,i)$
represents the annihilation operator of an electron with spin
$\ss$ in the $i$-th layer of the system. We assume that the
electronic spectrum in the layered system has the form
\be\label{energy}
\ve(\bk)=\ve(\bk_\parallel)+J\cos{(k_zs)}-E_F\;,
\ee
where $\bk\equiv(\bk_\parallel,k_z)$,
$\bk_\parallel\equiv(k_x,k_y)$, and $J$ is the effective hopping
energy between two adjacent layers situated at a distance $s$. The
attractive interaction leading to superconductivity,
$V(\bk,\bk')$, is assumed to be separable
\be\label{potentialBCS}
V(\bk,\bk')=|g|f(\bk)f(\bk')\;,
\ee
with $f(\bk)=[\cos(k_xa)-\cos(k_ya)]h_d(k_zs)$ for the case of
$d$-wave symmetry, $h_d(k_zs)$ being a function reflecting the $z$
axis dispersion of the system. The life time of the quasiparticle,
$\tau$, is introduce assuming the Born approximation for the
scattering potential and that the random potenatial obeys the
Gaussian ensemble
\bea\label{potential}
&&\overline{u_i(\br_\parallel)}=0\nonumber\\
&&\overline{u_i(\br_\parallel)u_j(\br'_\parallel)}=\f{1}{2\pi
N(0)\tau}\d_{ij}\d(\br_\parallel-\br'_\parallel)\;,
\eea
where the overline denotes the random average and $N(0)$ the
density of states at the Fermi surface. Under such assumptions the
quasiparticle Green's function is given by
\be\label{greenf}
G(\bk,i\o_n)=\f{1}{i\o_n\left[1+\left(2\tau|\o_n|\right)^{-1}\right]-\ve(\bk)}\;,
\ee
where $\o_n=(2n+1)\pi T$ is the standard fermionic Matsubara
frequency.

Following the original procedure introduced by Aslamazov and
Larkin\cite{larkin} the fluctuation propagator in the presence of
disorder can be calculated as
\be\label{fluctdef}
K^{-1}(\bq,i\o_n)=\f{1}{g N(0)}-\Pi(\bq,i\o_n)\;,
\ee
where in this definition $\o_n=2n\pi T$ denotes a bosonic
Matsubara frequency and
\be\label{pi}
\Pi(\bq,i\o_n)=T\sum_{\o_m}\sum_{\bk}|f(\bk)|^2G(\bk,i\o_m)G(\bq-\bk,i\o_n-i\o_m)\;.
\ee
Note that the summation over momenta $\bk$ in Eq. ({\ref{pi}) has
to be done assuming a cylindrical symmetry of the Fermi surface to
ensure that the correct symmetry imposed by the layered structure
of the system is well considered. The calculation of the
fluctuation propagator is straightforward
\bea\label{propagator}
&&K(\bq,i\o_n)\nonumber\\
&&=\f{1}{N(0)}\f{1}{\mu_0(T,D)+\gamma|\o_n|+\xi_0^2q^2_\parallel+
4\left(\f{\xi_{c0}}{s}\right)^2\sin^2{\left(\f{q_zs}{2}\right)}}\;,\nonumber\\
\eea
where $\mu_0(T,D)=2\pi^2(T\tau)/3+(D-D_c)/D_c$ gives the distance
to the phase transition point ($D=1/\tau$ represents the disorder
variable and $D_c=1.76\; T_{c0}$, $T_{c0}$ being the
superconducting critical temperature in a clean system),
$\gamma\simeq \tau$, $\xi_0^2=l^2/2$ ($l=v_F \tau$),
$\xi^2_{c0}=(J\tau s)^2/2$. According to the Thouless criterion
the phase transition occurs at $K^{-1}(0,0)=0$, meaning that in
our case it can be driven both by disorder or temperature. At the
quantum critical point (T=0) the phase transition is induced by
disorder, $\mu_0(T=0, D=D_c)=0$. On the other hand, in the weak
disorder limit, $D<D_c$, the phase transition occurs at a finite
temperature such that $\mu_0(T=T_c,D)=0$. The layered structure of
the considered system is well reflected in the form of the
fluctuation propagator $K(\bq, i\o_n)$ as it is easy to see from
Eq. (\ref{propagator}). At large interlayer separation,
$s\ra\infty$, the last term in the denominator of the right hand
side of the equation becomes small and the fluctuation propagator
corresponding to a quasi-2D system is recovered. When the
interlayer separation is small, $s\ra 0$, it is easy to see that
the fluctuation propagator will have the form corresponding to the
3D case. The crossover between 2D and 3D is then possible as a
function of the interlayer separation, $s$.

The general form of the action describing the phase fluctuations
in the critical region of the phase transition can be obtained
following the standard procedure\cite{ambegaokar} to decouple the
standard BCS action as
\bea\label{action}
&&S_{eff}=\sum_{q}\phi^\dagger(q)K^{-1}(q)\phi(q)\nonumber\\
&&+\f{u}{4}\sum_{q_1\cdots
q_4}\phi(q_1)\cdots\phi(q_4)\d\left(q_1+q_2+q_3+q_4\right)\;,
\eea
where $\phi(q)$ are the fluctuation field operators, and $u$
measure the interaction between fluctuations. In the above
equation, $q\equiv(\bq,\o_n)$ and
\bd
\sum_{q}\cdots\ra k_BT\sum_n\int\f{d^dq}{(2\pi)^d}\;,
\ed
$d$ being the system dimensionality.

\section{Renormalization-Group analysis of the transport
properties}

For a detailed analysis of the transport properties in the
critical region of a QPT we will use the general formalism of the
renormalization group approach introduced by Hertz\cite{hertz} and
developed lately by Millis.\cite{millis} The main idea in the
transport properties evaluation is that according to the
renormalization group approach the conductivity obeys the scaling
relation\cite{wang,phillips2}
\be\label{condscaling}
\ss_{\a\b}(\mu_0, T, \o, u)=e^{(d-2)l^*}\ss^*_{\a\b}\left[T(l^*),
\o(l^*), u(l^*)\right]\;,
\ee
where the scaling $l^*$ is defined such as the renormalization
procedure stop at $l^*$ given by $\mu(l^*)=1$. The asterisk
denotes that the conductivity is considered at the fixed point,
i.e., for $l=l^*$. For the most general case the renormalization
group equations corresponding to the action given by Eq.
(\ref{action}) can be obtained performing the standard scaling
$k=k'/b$ and $\o_n=\o'_n/b^z$ ($b=\ln{(l)}$) as
\begin{subequations}\label{ren}
\be\label{rengamma}
\f{d\Gamma(l)}{dl}=(z-2)\Gamma(l)\;,
\ee
\be\label{renT}
\f{dT(l)}{dl}=z\; T(l)\;,
\ee
\be\label{renmu}
\f{d\mu(l)}{dl}=2\mu(l)+\f{K_d \Gamma(l)
u(l)}{\exp{\left[\f{\Gamma(l)}{T(l)}\left(\Lambda^2+\mu(l)\right)\right]}-1}\;,
\ee
\bea\label{renu}
\f{du(l)}{dl}&=&\left[4-(d+z)\right]u(l)\nonumber\\
&-&\f{2K_d \Lambda^d T^2(l)}{4
T(l)\sinh^2{\left[\f{\Gamma(l)}{T(l)}\left(\Lambda^2+\mu(l)\right)\right]}}
u^2(l)\;,
\eea
\end{subequations}
where $z$ denotes the dynamical critical exponent,
$\Gamma=1/\gamma$, $K_d$ is a dimension dependent constant, and
$\Lambda(l)$ is a momenta cutoff. The set of Eqs. (\ref{ren})
admit an unstable Gaussian fixed point at $\Gamma=T=\mu=u=0$. In
the following we will consider separately the quantum disorder and
quantum critical regimes, as the renormalization group equations
lead to different solutions near the Gaussian fixed point.

\subsection{Quantum disorder regime}

In the quantum disorder regime ($\mu_0\gg T$) for the case $d=3$
and $z=2$ the renormalization group equations can be solved
relatively easy. The first two equations, (\ref{rengamma}) and
(\ref{renT}), lead to simple solutions, namely $\Gamma(l)=const.$
and $T(l)=T\;e^{2l}$, respectively. Because $d+z>4$, $u$ is
irrelevant and as a consequence the second term in Eq.
(\ref{renmu}) can be neglected leading to $\mu(l)=\mu_0\;e^{2l}$.
Accordingly, the renormalization procedure will be stopped at
\be\label{lstarQD}
l^*=\f{1}{2}\ln{\f{1}{\mu_0}}\;,
\ee
$\mu_0$ being the initial value of the distance to the phase
transition point. The system temperature at the fixed point
becomes
\be\label{temprenqd2}
T(l^*)=\f{T}{\mu_0}\;.
\ee
Note that a solution for the interaction term can be obtained as
$u(l^*)\sim \sqrt{\mu_0}$, a result which is in agreement with the
initial discussion of the interaction irrelevance in the $d=3$,
$z=2$ case.

\subsection{Quantum critical regime}

The quantum critical regime is defined by $\mu_0\ll T$. In this
case the integration of the renormalization group equations is
more complicated as the critical region around the phase
transition consists of two different domains, associated to
quantum and classical effects. The crossover between the two
domains is characterized by $\tilde{l}=[\ln{(1/T)}]/2$, such that
$T(\tilde{l})=1$. The integration over the scaling has to be split
in two domains, corresponding to quantum ($l<\tilde{l}$) and
classical ($\tilde{l}<l<l^*$) behavior. Consider now Eq.
(\ref{renmu}), which will lead to the general equation for the
scaling $l^*$ at which the renormalization procedure is stopped
($\mu(l^*)=1$). An approximate solution of this equation can be
obtained in two steps. First we introduce a new scaling variable,
$l'=l-\tilde{l}$, and secondly we neglect $\mu(l)$ in the
exponential term occurring in the right hand side of the equation.
The equation can be rewritten as
\be\label{renmuqc3D}
\f{d\mu(l')}{dl'}=2\mu(l')+\f{K_3\;\;u}{\exp{\left[e^{-2l'}\right]}-1}\;,
\ee
and admits the following solution
\bea\label{renmusolqc3D}
\mu(l')&=&\f{K_3\;u}{2}\ln{\left(\f{e-1}{\exp{\left[e^{-2l'}\right]-1}}\right)}\nonumber\\
&&-\f{K_3\;u}{2}\left(1-e^{-2l'}\right)\;.
\eea
Without loss of generality we assumed in both Eqs.
(\ref{renmuqc3D}) and (\ref{renmusolqc3D}) that $\Lambda^2=1$. The
renormalization procedure will be stopped at $l^*$ satisfying the
following condition
\be\label{ecl*}
\f{2}{K_3
u}=1-e^{2(l^*-\tilde{l})}+2(l^*-\tilde{l})e^{2(l^*-\tilde{l})}\;.
\ee
Finding an exact analytical solution for the above equation is not
possible, so we chose to solve the equation iteratively, the
solution within double logarithmic accuracy being of the form
\be\label{sol}
l^*=\f{1}{2}\ln{\left[\f{2}{T\;K_3\;u\;\ln{[2/(K_3\;u)]}}\right]}\;.
\ee
The corresponding renormalized temperature is
\be\label{rentempqc3D}
T(l^*)=\f{2}{K_3\;u\;\ln{[2/(K_3\;u)]}}\;.
\ee

In the following we will turn our attention to the system's
conductivity. We will evaluate the main contribution to the
conductivity in the framework of the renormalization group
approach.

\section{Fluctuation conductivity in layered systems}

The main contribution to the system conductivity will be
calculated following Aslamazov and Larkin\cite{larkin} based on
the Kubo formula. The layered structure of the system is
associated to the conductivity anisotropy in the system and
accordingly we will have to estimate different contributions for
the in-plane and $c$-axis conductivities. Following Varlamov {\it
et al.}\cite{varlamov} the main contribution to the conductivity
tensor can be calculated as
\be\label{condtens}
\ss^*_{\a\b}=-\lim_{\o\ra 0}\f{1}{i\o}[Q_{\a\b}]^R(\o)\;,
\ee
where $Q_{\a\b}(\o)$ represents the electromagnetic response
operator which contribute to the fluctuation conductivity of the
layered system. In Eq. (\ref{condtens}) the subscripts $(\a,\b)$
represents the polarization directions and $R$ denotes the
retarded part of the operator. A diagrammatic evaluation of the
electromagnetic response operator (see Ref. \onlinecite{varlamov})
leads to the following general form of the conductivity
\bea\label{condgen}
&&\ss^*_{\a\b}=\f{2\pi\xi_0^4m^2}{R_Q
T(l^*)}\int\f{d\o}{\sinh^2{\left(\f{\o}{2T(l^*)}\right)}}\nonumber\\
&&\times\int\f{d^2q_\parallel}{(2\pi)^2}
\int^{\pi/s}_{-\pi/s}\f{dq_z}{2\pi} v_\a v_\b \left[Im K^R(\bq,
\o)\right]^2\;,
\eea
where $R_Q=\pi\hbar/(2e^2)$, and $v_\a=[\6\ve(p)/\6p_\a]$. In Eq.
(\ref{condgen}) the initial integration over momenta was
considered based on the cylindrical symmetry of the system.
Accordingly, the in-plane and respectively the $c$-axis
conductivities become
\begin{widetext}
\be\label{inplanecond}
\ss^*_\parallel=\f{8\pi\gamma\xi_0^4 T(l^*)}{R_Q\mu_0^3}\int
d\o\;f\left(\f{\mu_0\o}{2\gamma
T(l^*)}\right)\int\f{d^2q_\parallel}{(2\pi)^2}
\int_{-\pi/s}^{\pi/s}\f{dq_z}{2\pi}\f{q^2_\parallel}
{\left\{\left[1+\f{\xi_0^2}{\mu_0}q^2_\parallel+
\f{4}{\mu_0}\left(\f{\xi_{c0}}{s}\right)^2\sin^2{\left(\f{q_z
s}{2}\right)}\right]^2+\o^2\right\}^2}
\ee
and
\be\label{caxiscond}
\ss^*_c=\f{8\pi\gamma\xi_0^4 T(l^*)}{R_Q\mu_0^3}\int
d\o\;f\left(\f{\mu_0\o}{2\gamma
T(l^*)}\right)\int\f{d^2q_\parallel}{(2\pi)^2}
\int_{-\pi/s}^{\pi/s}\f{dq_z}{2\pi}\f{m^2J^2s^2\sin^2{(q_z s)}}
{\left\{\left[1+\f{\xi_0^2}{\mu_0}q^2_\parallel+
\f{4}{\mu_0}\left(\f{\xi_{c0}}{s}\right)^2\sin^2{\left(\f{q_z
s}{2}\right)}\right]^2+\o^2\right\}^2}\;,
\ee
\end{widetext}
where $f(x)=x^2/[\sinh^2{(x)}]$. A similar way to investigate the
system conductivity was done in Ref. \onlinecite{adachi}, with the
specification that the final result for the in-plane conductivity
is obtained replacing the original layered symmetry of the system
with an isotropic 3D one. However, our approach is different, as
the original cylindrical symmetry of the system is conserved in
the conductivity calculation. The analytical structure of the
integrand in both Eqs. (\ref{inplanecond}) and (\ref{caxiscond})
allows us to distinguish between the QD and QC regimes.

\subsection{Quantum disorder regime}

In the quantum disorder regime, $\mu_0\gg T$, the main
contribution in the integration over the frequency variable in
both parallel and $c$-axis conductivities is associated to the
$\o=0$ point. Based on this approximation the two conductivities
can be calculated as
\be\label{inplanedisorder}
\ss^*_\parallel=\f{2\pi^2\gamma^2T^2(l^*)}{9R_Qs\mu_0^2}
\f{1+\f{2}{\mu_0}\left(\f{\xi_{c0}}{s}\right)^2}
{\left\{\left[1+\f{2}{\mu_0}\left(\f{\xi_{c0}}{s}\right)^2\right]^4-
\f{4}{\mu_0^2}\left(\f{\xi_{c0}}{s}\right)^4\right\}^{3/2}}\;,
\ee
and
\be\label{caxisdisorder}
\ss^*_c=\f{2\pi^2\gamma^2T^2(l^*)}{9R_Q s\mu_0^3}
\f{s^2\xi_0^2m^2J^2}
{\left\{\left[1+\f{2}{\mu_0}\left(\f{\xi_{c0}}{s}\right)^2\right]^4-
\f{4}{\mu_0^2}\left(\f{\xi_{c0}}{s}\right)^4\right\}^{3/2}}\;.
\ee
A better understanding of the in-plane and $c$-axis conductivities
is achieved in the limit of large and small separation between
adjacent layers of the system.

\subsubsection{Large interlayer separation, $s\ra\infty$}

For the large interlayer separation case, $s\ra\infty$, which from
the dimensional point of view approaches a quasi 2D system, the
system's conductivities evaluated at the fix point are
\be\label{sp2DQD}
\ss_\parallel=\f{2\pi^2\gamma^2_1}{9R_Qs}\f{T^2}{\mu_0^6}
\ee
and
\be\label{sc2DQD}
\ss_c=\f{2\pi^2\gamma_1^2\xi_0^2m^2J^2s}{9R_Q}\f{T^2}{\mu_0^7}\;.
\ee
Note that Eqs. (\ref{sc2DQD}) and (\ref{sp2DQD}) were obtained
using the renormalized values for the system's temperature in the
QD regime. The in-plane and $c$-axis conductivities have both the
same temperature dependence but different disorder dependence
close to the QPT point.

\subsubsection{Small interlayer separation, $s\ra 0$}

In the case of small interlayer separation, $s\ra 0$, the system
approaches a quasi 3D system. A simple calculation of the system's
conductivities at the fixed point leads to
\be\label{sp3DQD}
\ss_\parallel=\f{\pi^2\gamma^2_1}{596
R_Q\xi_{c0}}\f{T^2}{\mu^{11/2}_0}
\ee
and
\be\label{sq3DQD}
\ss_c=\f{\pi^2\gamma^2_1\xi^2_0m^2J^2s^4}{36 R_Q
\xi^3_{c0}}\f{T^2}{\mu^{11/2}_0}\;.
\ee
Different from the large interlayer separation case, in the
quasi-3D case the temperature and disorder dependence of the two
conductivities is the same.

\subsection{Quantum critical regime}

In the QC regime, $\mu_0\ll T$, we approximate $f(x)\ra 1$, as in
this situation $x\ll 1$. The analytical forms of the in-plane and
$c$-axis conductivities are obtained as
\be\label{inplanecritical}
\ss_\parallel=\f{\pi\gamma T}{2R_Q s\mu_0}\f{1}
{\left\{\left[1+\f{2}{\mu_0}\left(\f{\xi_{c0}}{s}\right)^2\right]^4-
\f{4}{\mu_0^2}\left(\f{\xi_{c0}}{s}\right)^4\right\}^{1/2}}
\ee
and
\begin{widetext}
\bea\label{caxiscritical}
\ss_c=\f{\pi\gamma T\xi_0^2m^2J^2s}{2R_Q\mu_0^2}
\f{1+\f{2}{\mu_0}\left(\f{\xi_{c0}}{s}\right)^2-\left\{\left[1+\f{2}{\mu_0}\left(\f{\xi_{c0}}{s}\right)^2\right]^4-
\f{4}{\mu_0^2}\left(\f{\xi_{c0}}{s}\right)^4\right\}^{1/2}}
{\f{4}{\mu_0^2}\left(\f{\xi_{c0}}{s}\right)^4\left\{\left[1+\f{2}{\mu_0}\left(\f{\xi_{c0}}{s}\right)^2\right]^4-
\f{4}{\mu_0^2}\left(\f{\xi_{c0}}{s}\right)^4\right\}^{1/2}}\;.
\eea
\end{widetext}
In the following we will consider a simplified form of the two
obtained equations in the limit of large and small interlayer
separation in order to discuss the temperature and disorder
dependence of the conductivity.

\subsubsection{Large interlayer separation, $s\ra\infty$}

For the large interlayer separation between two adjacent planes of
the layered structure the in-plane and $c$-axis conductivities can
be approximated as
\be\label{sp2DQC}
\ss_\parallel=\f{4\pi\gamma_1}{R_Q s \left(K_3u\ln{\f{2}{K_3
u}}\right)^3}\f{1}{\mu_0 T^2}
\ee
and
\be\label{sc2DQC}
\ss_c=\f{2\pi\gamma_1\xi^2_om^2J^2s}{R_Q\left(K_3u\ln{\f{2}{K_3
u}}\right)^3}\f{1}{\mu^2_0 T^2}\;.
\ee
As in the QD regime, the temperature dependence of the two
conductivities at large separation between adjacent layers of the
system is the same, however, the disorder dependence of the two
conductivities differs.

\subsubsection{Small interlayer separation, $s\ra 0$}

Let us consider now the small interlayer separation case,
resembling the quasi 3D-system. The in-plane and $c$-axis
conductivities are obtained as
\be\label{sp3DQC}
\ss_\parallel=\f{2\pi\gamma_1}{R_Q\xi_{c0}\left(K_3u\ln{\f{2}{K_3
u}}\right)^3}\f{1}{\mu^{1/2}_0 T^2}
\ee
and
\be\label{sc3DQC}
\ss_c=\f{\pi\gamma_1\xi^2_0m^2J^2s^4}{R_Q
\xi^3_{c0}\left(K_3u\ln{\f{2}{K_3 u}}\right)^3}\f{1}{\mu^{1/2}_0
T^2}\;.
\ee
Once again for the quasi-3D case the temperature and disorder
dependence of the two conductivities is similar.

\section{Conclusions}

In conclusion we presented a detailed analysis of the in-plane and
$c$-axis conductivities close to the QPT points identified in
$d$-wave superconductors. We considered that the system is well
described by a layered structure of coupled planes, a model which
allowed us to analytically obtained the two components of the
system's conductivity. For our model, the system dimensionality is
fixed at $d=3$ despite the fact that large interlayer separation
resemble a quasi-2D dimensional system; at small interlayer
separation the anisotropic 3D system is recovered. The system's
dimensionality, $d=3$, together with the dynamical critical
exponent, $z=2$, makes the results of our calculation suitable to
describe the QPT in the overdoped region of HTSC phase diagram.
However, the possibility to interpolate between quasi-2D and 3D
systems due to the layered structure of the system can provide a
better understanding of the disorder induced phase transition in
HTSC in both underdoped and overdoped regions of the phase
diagram.

Our estimation of the system's conductivity was based on the
renoprmalization group approach. The analysis of the possible
QPT's in $d$-wave superconductors was done by
Schneider\cite{schneider1,schneider} based on a detailed revision
of experimental data for correlation length, magnetic penetration
depth, specific heat and resistivity. His conclusions are
different for the two QPT points observed in the underdoped and
overdoped region of the phase diagram. In the underdoped region a
quantum superconductor-insulator transition was identified, whose
characteristics are $d=2$ and $z=1$. On the other hand, in the
overdoped region a superconductor-normal state transition was
identified with $d=3$ and $z=2$. The choice of our system symmetry
and implicitly the form of our fluctuation propagator forced us to
consider a fix dimensionality of the system, namely $d=3$, meaning
that the point d=2 is not accessible to our approximation. We also
considered $z=2$ according to the experimental data. However, as
we can vary the interlayer separation distance we were able to
approach a quasi-2D system and show that a possible QPT is present
even in this situation, both as function of disorder or
temperature. In the $d=3$, $z=2$ limit the analysis of the
renormalization group equations is straightforward, leading to a
Gaussian fixed point. In this situation the quadratic term in the
system action, describing direct interaction between fluctuations,
is irrelevant, meaning that for a certain range of temperature and
disorder around the fixed point we can calculate the systems's
conductivity in the perturbative renormalization group.

We considered two different regimes in our approximations, namely
the quantum disorder regime, $T\ll\mu_0$, and the quantum critical
regime, $T\gg\mu_0$. Analytical results were presented for both
type of conductivities in two different limits, for large and
small separation between the component layers of the system. As a
general rule the temperature dependence of the these
conductivities is similar for all situations in our calculation,
as long as we consider the same regime. On the other hand, for
both large and small interlayer separation limits, when we lower
the temperature the system resistivity should be nonmonotonic, a
dip at a temperature of the order $T\sim \mu_0$ being present. A
similar result was reported in Ref. \onlinecite{phillips2} for the
case $d=2$ and $z=2$, which actually can be seen as a extreme
limit of our calculation. The situation changes as we investigate
the disorder dependence of the conductivities. At small separation
between adjacent layers, when we approach a quasi-3D system both
in-plane and $c$-axis conductivities diverge at the critical
disorder point following the same power law, whatever we are in
the quantum disorder or quantum critical regime. At large
separation, when the system resemble a quasi-2D one, the behavior
of the in-plane and $c$-axis conductivities at the QCP are
different as function of disorder. Despite the fact that both
conductivities diverge at the critical disorder point following
power laws, the divergence observed for the $c$-axis conductivity
is stronger.

The case of a pure 2D system was previously considered by
Dalidovich and Phillips\cite{phillips2} at finite temperature, and
by Herbut at $T=0$.\cite{herbut} Our result for the in-plane
conductivity, at large separation between adjacent layers of the
system, is similar to the one obtained in Ref.
\onlinecite{phillips2}. Some differences occurs due to the
different system dimensionality and a difference between the form
of the quasiparticle propagator consider here and in Ref.
\onlinecite{phillips2}. However, the main result for the two
dimensional case, which states that the in-plane conductivity in
the quantum critical regime is non universal and increase as
function of temperature when the QCP ($T\ra 0$) is approached was
also proved in our calculation. Moreover, our investigation
presents qualitative results also for the quasi-3D case, which is
relevant for the case of $d$-wave superconductors, where, despite
the fact that conduction in the superconducting phase is
attributed to the CuO planes, the system is a 3D one.

\begin{acknowledgments}
One of the authors (MC) would like to acknowledge stimulating
discussions with A. A. Varlamov, P. Phillips, and A.-M. Tremblay.
We acknowledge financial support from the Romanian Ministry of
Education.
\end{acknowledgments}

\end{document}